# Towards full control of molecular exciton energy transfer via FRET in DNA origami assemblies

*Aleksandra K. Adamczyk[1], Teun A.P.M. Huijben[2], Karol Kołątaj[1], Fangjia Zhu[1], Rodolphe Marie[2], Fernando D. Stefani[3,4], and Guillermo P. Acuna[1]\**

[1] Department of Physics, University of Fribourg, Chemin du Musée 3, Fribourg CH-1700, Switzerland.

[2] Department of Health Technology, Technical University of Denmark, Anker Engelunds Vej 101, 2800 Kongens Lyngby, Denmark.

[3] Centro de Investigaciones en Bionanociencias (CIBION), Consejo Nacional de Investigaciones Científicas y Técnicas (CONICET), Godoy Cruz 2390, C1425FQD Ciudad Autónoma de Buenos Aires, Argentina.

[4] Departamento de Física, Facultad de Ciencias Exactas y Naturales, Universidad de Buenos Aires, Güiraldes 2620, C1428EHA Ciudad Autónoma de Buenos Aires, Argentina.

**Corresponding Authors**

\* E-mail: guillermo.acuna@unifr.ch (G.P.A.)

**Keywords:** DNA nanotechnology, nanofabrication, DNA origami, single-molecule, FRET, fluorescence.

**Abstract**

Controlling the flow of excitons between organic molecules holds immense promise for various applications, including energy conversion, spectroscopy, photocatalysis, sensing, and microscopy. DNA nanotechnology has shown promise in achieving this control by using synthetic DNA as a platform for positioning and, very recently, for also orienting organic dyes. In this study, the orientation of doubly-linked dyes in DNA origami structures was manipulated to control energy transfer. By controlling independently the orientation of single donor and acceptor molecules, the average energy transfer efficiency was doubled. This work demonstrates the potential of DNA nanotechnology for precise control of the excitonic energy transfer with implications for artificial light-harvesting antennas.

**Main Text**

The ability to control the flow of excitons between organic molecules, is a long-standing goal due to its multiple implications for spectroscopy, photocatalysis, energy transfer and conversion, sensing, and microscopy[1]. Furthermore, such level of excitonic control could be vital not only for the fabrication of artificial light-harvesting complexes[2] but also for the conception of complex collective effects such as superradiance and superfluorescence for the next generation of laser sources[3]. While tremendous progress was achieved over the past decades in understanding and tailoring energy transfer among molecules and other entities, artificial systems are still very far from the benchmark achieved by nature over millions of years of evolution as photosynthetic systems can transport energy to the reaction centers with a quantum efficiency close to unity[4,5]. This is achieved through a combination of processes in which Förster (fluorescence) resonant energy transfer (FRET) plays the main role[6]. A major breakthrough in our attempts to mimic nature´s unmatched capacity to transfer energy between molecules with minor losses was enabled by DNA nanotechnology[7]. Synthetic DNA is programmable, flexible in terms of size and design, currently widely available at a low cost, and can be modified to host a wide variety of organic fluorophores covalently attached or directly intercalated (non-covalent bond) within the double-stranded (ds) DNA helix[8–10]. These unique advantages were exploited in several studies of FRET between single molecules where both the position and stoichiometry of organic dyes could be controlled throughout a ds-DNA helix[11–17]. The advent of more complex DNA assemblies and in particular the DNA origami technique[18,19] lead to a myriad of elaborated arrangements of organic dyes in two and three dimensions that can more closely resemble those occurring in natural light-harvesting complexes[20–30]. These works clearly consolidate the DNA origami technique as an ideal platform[31] to position organic fluorophores and even natural light-harvesting complexes[32] with nanometer precision and stoichiometric control to study exciton energy transfer via FRET.

FRET processes are described by a dipole-dipole interaction between a donor-acceptor pair which is strongly distant dependent with an energy transfer rate proportional to $d^{-6}$ (with $d$ the intermolecular separation distance). In addition, FRET is also critically dependent on the relative orientation between the transition dipole moment of the molecules involved in the energy transfer. Formally, this dependency is described by the factor $\kappa^2$, that can take values between 0 and 4 depending on the relative orientation of the donor and acceptor dipole moments[33], see **Figure 1a**. To date, DNA nanotechnology has led to an unprecedented nanometric control over the intermolecular distance $d$ in synthetic assemblies. However, controlling the orientation of dyes in DNA assemblies remains challenging as it depends on a multiplicity of factors including binding strategy, local environment, and fluorophore identity[34]. For intercalating dyes (non-covalent attachment), the orientation is typically unambiguously determined by the intercalating strategy. However, this occurs at the expense of neither positioning nor stoichiometric control, as it is not possible to predefine the number or the positions of the intercalating molecules[35]. The orientation of dyes covalently attached to ds-DNA has been the matter of study of a few works[36,37]. Additionally, it was found that Cy3-Cy5 (donor-acceptor) FRET pairs terminally attached to the 5´ ends of ds-DNA can stack in a defined orientation[13]. In this way, their relative orientation can be tuned, although it is entangled with the donor-acceptor separation distance and the two parameters cannot be controlled independently[8,14]. Recently, we have studied the orientation and wobbling of different dyes in DNA origami structures using a combination of several imaging techniques[38–40]. We and others have found that the orientation and wobbling of Cy3 and Cy5 fluorophores doubly-linked to single-stranded (ss) DNA in DNA origami constructs can be controlled by engineering the number of adjacent paired and unpaired nucleotides. In particular, we have demonstrated that for both Cy3 and Cy5 dyes, their in-plane molecular orientation can be switched from being parallel to perpendicular to the dsDNA helix-bundle in the origami host[41–43]. In this work, we study how this technique can be used to manipulate the energy transfer between single Cy3-Cy5 donor-acceptor pairs doubly-linked to DNA origami at a fixed separation distance of 6 nm with two different relative orientations. We performed FRET measurements on single surface-immobilized structures, thus avoiding typical issues of ensemble measurements, such as variations in FRET signal due to different donor or acceptor concentrations. We obtain the FRET efficiency, $E$, using the acceptor bleaching approach[44] based solely on the donor´s intensity before and after the acceptor bleaching[45]. In this way, we get an accurate estimation of $E$ without the need for corrections on the dyes´ quantum yield and detection efficiency at different wavelengths[46]. Our results show that the average value of $E$ can be doubled by solely controlling the relative orientation between the donor-acceptor pair. These findings are in agreement with FRET calculations based on the previously reported angular distribution of the dyes and correspond to $\kappa^2$ values of 0.32 and 0.67 for the two different configurations. This work represents the first step towards achieving full control of molecular excitonic energy transfer and provides novel tools for the development of artificial light-harvesting antennas that can better mimic nature.

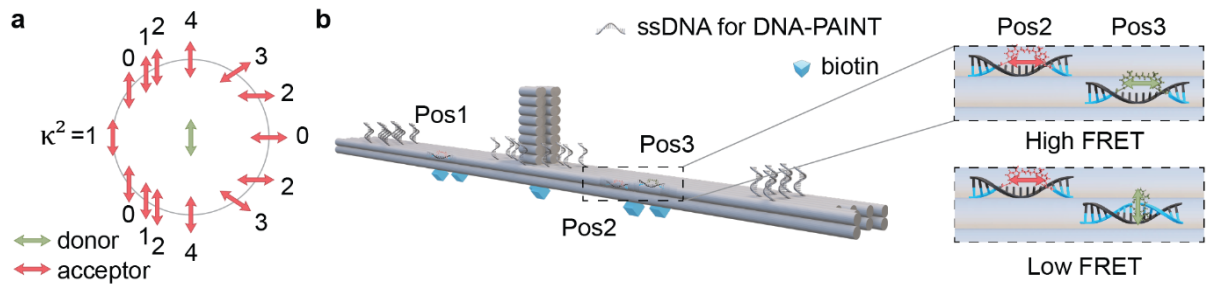

***Figure 1**. a) $\kappa^2$ factor for different in-plane relative orientations of the transition dipole moments of the donor (green, center) and the acceptor (red, periphery). $\kappa^2$ can take values between 0 and 4. b) DNA origami structure including three sites with ssDNA extensions for DNA PAINT, biotin modifications for immobilization on the substrate, and three positions (Pos1, Pos2 and Pos3) where the fluorophores were incorporated (left). Zoom at positions 2 and 3, where a single donor-acceptor pair is placed at the same distance but with different orientations (indicated by the red/green arrows) to achieve "High FRET" and "Low FRET" (right).*

**Figure 1b** depicts the DNA origami host structure employed (more details in[47], specific modifications can be found in the supporting information). The structure is modified with biotins at the bottom for surface immobilization and with three groups of ssDNA extensions for DNA PAINT super-resolution measurements[40]. For the FRET studies, we incorporate doubly-linked Cy3 and Cy5 dyes into the DNA origami. We employ the technique developed recently[42,43] to engineer the environment of each dye in order to control the in-plane orientation. In particular, we leave either eight or no neighboring nucleotides unpaired. In this way, the dye can be "stretched" and aligned parallel to the host DNA double-helix or remain perpendicular to it[43]. We fabricate three different samples labelled "No FRET", "Low FRET" and "High FRET". Samples "Low FRET" and "High FRET" include a single FRET pair, with a Cy3 (donor) at Pos3 and a Cy5 (acceptor) at Pos2, separated by a 6 nm distance. The only difference between these samples lies in the donor orientation, see zoom in **figure 1b**. The High FRET (Low FRET) sample is based on a Cy3 ∥-Cy5 ∥ (Cy3 ⊥- Cy5 ∥) donor-acceptor pair. "No FRET" is a control reference structure with a Cy5 at Pos1 together with a Cy3 at Pos3. In this sample, the donor and the acceptor are separated by 50 nm, and thus no FRET is expected regardless of the relative orientation between the dyes.

First, we confirmed that the dyes were oriented according to the sample design through a combination of DNA PAINT super-resolution and polarization-resolved excitation measurements[41] on surfaced immobilized samples. In addition, we employed PCA/PCD/Trolox buffer for the photostabilization of the Cy3 and Cy5 dyes. For these experiments, DNA origami structures with a single dye were self-assembled, i.e. containing either the acceptor or the donor at each orientation. **Figure 2a** includes the spherical coordinate system employed. This combination of measurements can retrieve the in-plane angle $\rho$ and also an estimate of the dye's "wobbling" through the modulation $M$, where $M = 1$ and $M = 0$ corresponds to a fixed dye and to a freely rotating dye (in a time scale faster than the polarization rotation, see supporting information) respectively. The measured orientations and modulations for the Cy5 ∥ (Pos2), Cy3 ∥ and Cy3 ⊥ (Pos3) are included in **figures 2b, 2c** and **2d** respectively. These results are in line with previous reports[42,43], and further consolidate the fact that the orientation of doubly-linked Cy3 and Cy5 dyes can be controlled to align predominantly parallel or perpendicular to the host double-helix in DNA origami with a certain degree of wobbling, particularly for the parallel cases.

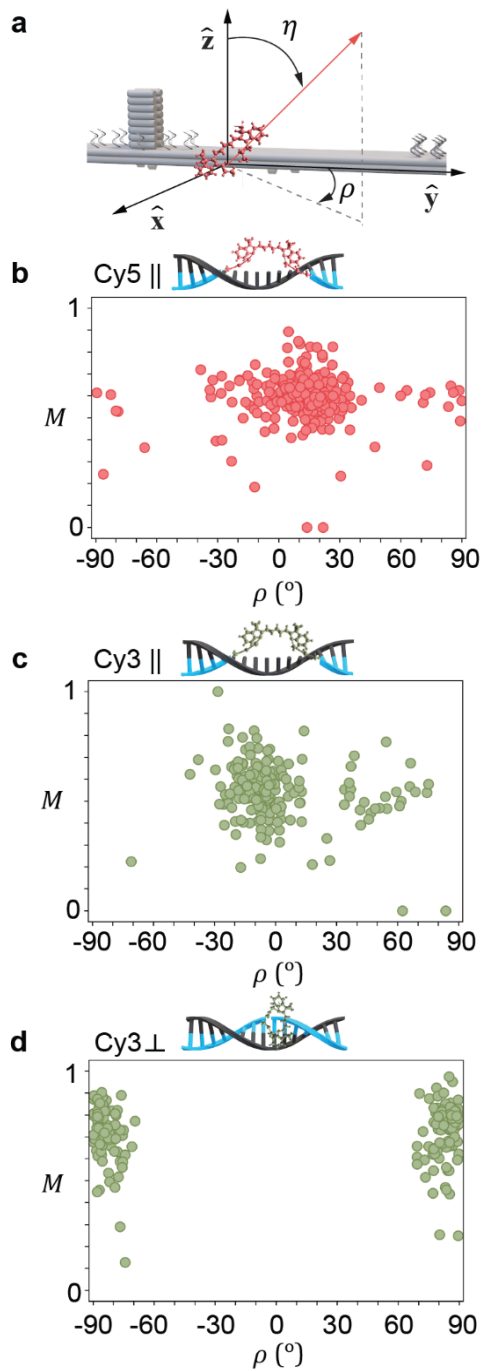

*Figure 2. a) The spherical coordinate system employed for the determination of the in-plane angle ρ and wobbling M. M vs ρ extracted from a combination of super-resolution and polarization-resolved excitation measurements, for b) the Cy5 ∥ at Pos2, c) Cy3 ∥ at Pos3, and d) Cy3 ⊥ at Pos3 d). In all case at least 190 single dyes were measured.*

Next, we performed the FRET measurements for the three different samples. The challenges in the reliable and non-arbitrary determination of the FRET efficiency, *E*, have been investigated in several works[16,48]. Here we avoid most of the uncertainties by following the acceptor bleaching approach at the single molecule level, alternating between donor and acceptor excitation while acquiring fluorescent transients in both the donor and the acceptor emission spectral ranges. **Figure 3a** includes a sketch of the measurement pipeline employed. In a wide-field fluorescence microscope, we excite first the surface immobilized samples with a 532 nm laser ("green" donor excitation). Then, we switch to an illumination at 640 nm ("red" acceptor excitation) until the acceptor dye bleaches. After that, we switch back to donor excitation until the donor dye bleaches. In this approach, the FRET efficiency is calculated as, $E = (I_d - I_{da})/I_d$ with $I_{da}$ and $I_d$ the donor´s intensity in the presence and absence of the acceptor respectively, thus using a single detection channel. Furthermore, since measurements are only taken into account for further analysis when the bleaching of both the donor and the acceptor is detected, the presence of a single Cy3-Cy5 pair is confirmed[45].

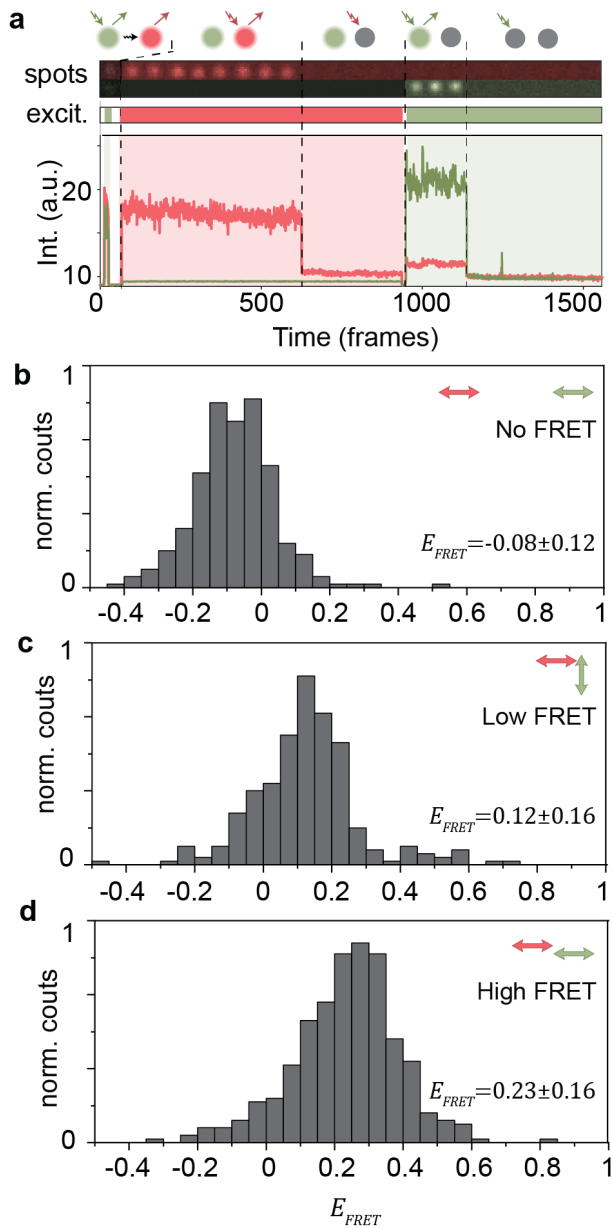

*Figure 3.* a) FRET measurement pipeline for the acceptor bleaching approach. First, the donor (green) is excited, and both the acceptor (red) and the donor are detected. Then only the acceptor is excited until it bleaches. Finally, the donor is excited again until it bleaches. While the excitation is alternating, fluorescence is detected in both spectral ranges and fluorescence transients are generated. FRET efficiency histogram plots for the three different samples, No FRET b), Low FRET c) and High FRET d). For each case, more than 250 single FRET pairs were measured.

**Figures 3b-d** include histograms of the FRET efficiency results obtained for the three different samples with more than 250 single donor-acceptor pairs each. As expected, the "No FRET" sample exhibits a marginal average FRET efficiency of $E_{NF} = -0.08 \pm 0.12$ (**Figure 3b**) which sets the overall uncertainty of our measurement approach. In contrast, the remaining two samples show a significantly higher FRET efficiency. For the "Low FRET" origami structures (**Figure 3c**), we measure an average $E_{LF} = 0.12 \pm 0.16$. As expected, the "High FRET" (**Figure 3d**) sample yields an even higher efficiency with an average value of $E_{HF} = 0.23 \pm 0.16$ which corresponds to approximately a ×2 increment. We emphasize that the difference in FRET efficiency between the Low FRET and High FRET samples is almost entirely due to the different orientations among the donor-acceptor pairs, as dyes are placed at the same relative position within the dsDNA turn of the double helix.

In order to rationalize these findings, we proceeded to calculate $E_{LF}$ and $E_{HF}$. For the two dyes used as donor-acceptor pair, we have characterized their in-plane angular distribution for the two orientations employed (∥ and ⊥, see **figure 2**). Still, the $\kappa^2$ factor depends on both the in-plane and the out-of-plane ($\eta$, **figure 2a**) relative orientations. In this work, the $\eta$ was not determined due to the challenging combination of measurements required[43]. However, based on the good agreement between the in-plane angular distributions displayed in **figure 2** for the ∥ and ⊥ cases and those obtained previously, here we use the out-of-plane angular distribution measured in[43] to estimate $\kappa^2$. The distribution of $\kappa^2$ for the "Low FRET" and "High FRET" samples are included in **figure 4** (further details on the $\kappa^2$ calculation can be found in the supporting information) with average values of $\kappa^2_{LF} = 0.32$ and $\kappa^2_{HF} = 0.67$. These values are in good qualitative agreement with the increment by a factor of two for the FRET efficiency between the Low and High FRET samples. It is worth mentioning that the dyes´ wobbling was not considered for the estimation of $\kappa^2$. Finally, based on the donor-acceptor separation of 6 nm, $\kappa^2$, and the spectral overlap integral[49] between Cy3 and Cy5, we calculated the expected

FRET efficiency, $E'$. For the "Low FRET" and "High FRET" samples we obtain values of $E'_{LF} = 0.154$ and $E'_{HF} = 0.278$ respectively in excellent agreement with our measurements.

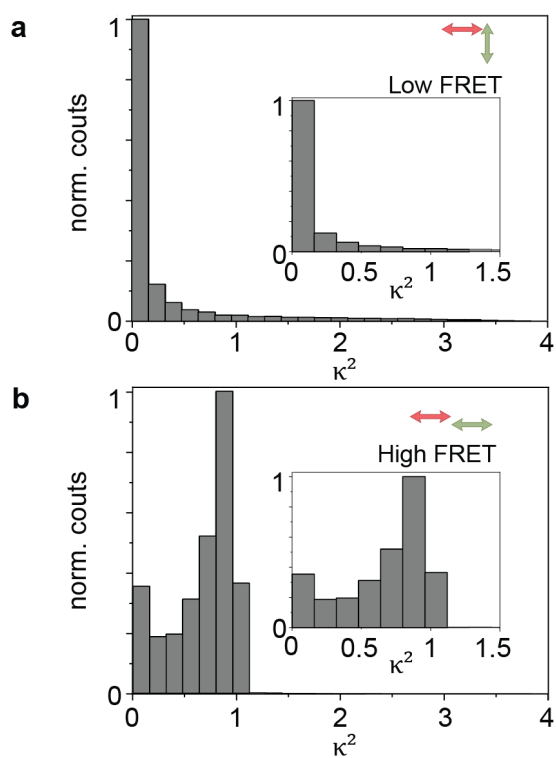

*Figure 4. Histogram plot of the calculated $\kappa^2$, for the Low FRET and High FRET samples, a) and b) respectively. In both cases, the inset depicts a zoom with $\kappa^2$ values between 0 and 1.5. The calculations were based on the in-plane angles determined in figure 2 combined with the out-of-plane angles measured previously[43] and led to average values of $\kappa^2_{LF} = 0.32$ and $\kappa^2_{HF} = 0.67$.*

**Discussion**

In summary, we have demonstrated that orientational control of doubly-linked fluorophores in DNA origami can be used to adjust exciton energy transfer between molecules via FRET. We show that the FRET efficiency of a single donor-acceptor pair can be enhanced by a factor of 2 by controlling the relative orientation of the donor and acceptor. Remarkably, these experimental results were rationalized with an excellent agreement with numerical calculations of the $\kappa^2$ and the FRET efficiency based on the independent measurements of the dyes´ orientations. This work advances the degree of control of DNA nanotechnology to organize single molecules, providing independent control of molecular position and orientation, which is vital for the manipulation of energy transfer. Further, the demonstrated manipulation of energy transfer between two molecules, can be considered as just the first step towards the fabrication of artificial light-harvesting devices where multiple fluorophores are organized in space to maximize energy funneling in predetermined directions.

**Author contributions**

F.D.S., and G.P.A. conceived the experiments. A.K.A, T.A.P.M.H., and G.P.A. designed the experiments. A.K.A., T.A.P.M.H., and K.K. designed and fabricated the DNA origami structures. A.K.A and T.A.P.M.H. performed the preliminary fluorescence measurements. T.A.P.M.H. wrote the FRET analysis software. A.K.A. performed final experiments and analyzed results. F.Z. performed $\kappa^2$ calculations. R.M., F.D.S., and G.P.A. supervised the project. A.K.A. made figures and supplementary information. G.P.A. wrote the manuscript with input from all of the authors.

**Acknowledgements**

This project has received funding from the European Union's Horizon 2020 research and innovation programme under the Marie Skłodowska-Curie grant agreement No 860914. F.D.S. acknowledges the support of the Max Planck Society and the Alexander von Humboldt Foundation. This work has been funded by Consejo Nacional de Investigaciones Científicas y Técnicas (CONICET) and Agencia Nacional de Promoción Científica y Tecnológica (ANPCYT), project PICT-2017-0870. G.P.A. acknowledges support from the Swiss National

Science Foundation (200021_184687) and through the National Center of Competence in Research Bio-Inspired Materials (NCCR, 51NF40_182881), the European Union Program HORIZON-Pathfinder-Open: 3D-BRICKS, grant Agreement 101099125. FDS acknowledges the Alexander von Humboldt foundation for their constant support.

**References**

(1) Mathur, D.; Díaz, S. A.; Hildebrandt, N.; Pensack, R. D.; Yurke, B.; Biaggne, A.; Li, L.; Melinger, J. S.; Ancona, M. G.; Knowlton, W. B.; et al. Pursuing Excitonic Energy Transfer with Programmable DNA-Based Optical Breadboards. *Chem. Soc. Rev.* **2023**, *52* (22), 7848–7948.

(2) Kundu, S.; Patra, A. Nanoscale Strategies for Light Harvesting. *Chem. Rev.* **2017**, *117* (2), 712–757.

(3) Holzinger, R.; Oh, S. A.; Reitz, M.; Ritsch, H.; Genes, C. Cooperative Subwavelength Molecular Quantum Emitter Arrays. *Phys. Rev. Res.* **2022**, *4*.

(4) Hogewoning, S. W.; Wientjes, E.; Douwstra, P.; Trouwborst, G.; van Ieperen, W.; Croce, R.; Harbinson, J. Photosynthetic Quantum Yield Dynamics: From Photosystems to Leaves. *Plant Cell* **2012**, *24* (5), 1921–1935.

(5) Genty, B.; Briantais, J. M.; Baker, N. R. The Relationship between the Quantum Yield of Photosynthetic Electron Transport and Quenching of Chlorophyll Fluorescence. *Biochim. Biophys. Acta - Gen. Subj.* **1989**, *990* (1), 87–92.

(6) Feron, K.; Belcher, W. J.; Fell, C. J.; Dastoor, P. C. Organic Solar Cells: Understanding the Role of Förster Resonance Energy Transfer. *Int. J. Mol. Sci. 2012, Vol. 13, Pages 17019-17047* **2012**, *13* (12), 17019–17047.

(7) Hart, S. M.; Gorman, J.; Bathe, M.; Schlau-Cohen, G. S. Engineering Exciton Dynamics with Synthetic DNA Scaffolds. *Acc. Chem. Res.* **2023**, *56* (15), 2051–2061.

(8) Armitage, B. A. Cyanine Dye--DNA Interactions: Intercalation, Groove Binding, and Aggregation. In *DNA Binders and Related Subjects: -/-*; Waring, M. J., Chaires, J. B., Eds.; Springer Berlin Heidelberg: Berlin, Heidelberg, 2005; pp 55–76.

(9) Stadler, A. L.; Delos Santos, J. O.; Stensrud, E. S.; Dembska, A.; Silva, G. L.; Liu, S.; Shank, N. I.; Kunttas-Tatli, E.; Sobers, C. J.; E Gramlich, P. M.; et al. Fluorescent DNA Nanotags Featuring Covalently Attached Intercalating Dyes: Synthesis, Antibody Conjugation, and Intracellular Imaging. **2011**.

(10) Ihmels, H.; Otto, D. Intercalation of Organic Dye Molecules into Double-Stranded DNA -- General Principles and Recent Developments. In *Supermolecular Dye Chemistry*; Springer-Verlag: Berlin/Heidelberg, 2005; pp 161–204.

(11) Didenko, V. V. DNA Probes Using Fluorescence Resonance Energy Transfer (FRET): Designs and Applications. *https://doi.org/10.2144/01315rv02* **2018**, *31* (5), 1106–1121.

(12) Dietrich, A.; Buschmann, V.; Müller, C.; Sauer, M. Fluorescence Resonance Energy Transfer (FRET) and Competing Processes in Donor–Acceptor Substituted DNA Strands: A Comparative Study of Ensemble and Single-Molecule Data. *Rev. Mol. Biotechnol.* **2002**, *82* (3), 211–231.

(13) Ouellet, J.; Schorr, S.; Iqbal, A.; Wilson, T. J.; Lilley, D. M. J. Orientation of Cyanine Fluorophores Terminally Attached to DNA via Long, Flexible Tethers. *Biophys. J.* **2011**.

(14) Iqbal, A.; Arslan, S.; Okumus, B.; Wilson, T. J.; Giraud, G.; Norman, D. G.; Ha, T.; Lilley, D. M. J. Orientation Dependence in Fluorescent Energy Transfer between Cy3 and Cy5 Terminally Attached to Double-Stranded Nucleic Acids. *Proc. Natl. Acad. Sci. U. S. A.* **2008**.

(15) Clegg, R. M.; Murchie, A. I. H.; Zechel, A.; Lilley, D. M. J. Observing the Helical Geometry of Double-Stranded DNA in Solution by Fluorescence Resonance Energy Transfer. *Proc. Natl. Acad. Sci.* **1993**, *90* (7), 2994–2998.

(16) Hellenkamp, B.; Schmid, S.; Doroshenko, O.; Opanasyuk, O.; Kühnemuth, R.; Rezaei Adariani, S.; Ambrose, B.; Aznauryan, M.; Barth, A.; Birkedal, V.; et al. Precision and Accuracy of Single-Molecule FRET Measurements—a Multi-Laboratory Benchmark Study. *Nat. Methods* **2018**, *15* (9), 669–676.

(17) Cunningham, P. D.; Khachatrian, A.; Buckhout-White, S.; Deschamps, J. R.; Goldman, E. R.; Medintz, I. L.; Melinger, J. S. Resonance Energy Transfer in DNA Duplexes Labeled with Localized Dyes. *J. Phys. Chem. B* **2014**, *118* (50), 14555–14565.

(18) Rothemund, P. W. K. Folding DNA to Create Nanoscale Shapes and Patterns. *Nature*. 2006, pp 297–302.

(19) Douglas, S. M.; Dietz, H.; Liedl, T.; Högberg, B.; Graf, F.; Shih, W. M. Self-Assembly of DNA into Nanoscale Three-Dimensional Shapes. *Nature* **2009**, *459* (7245), 414–418.

(20) Stein, I. H.; Steinhauer, C.; Tinnefeld, P. Single-Molecule Four-Color FRET Visualizes Energy-Transfer Paths on DNA Origami. *J. Am. Chem. Soc.* **2011**, *133* (12), 4193–4195.

(21) Mathur, D.; Samanta, A.; Ancona, M. G.; Díaz, S. A.; Kim, Y.; Melinger, J. S.; Goldman, E. R.; Sadowski, J. P.; Ong, L. L.; Yin, P.; et al. Understanding Förster Resonance Energy Transfer in the Sheet Regime with DNA Brick-Based Dye

Networks. *ACS Nano* **2021**, *15* (10), 16452–16468.

(22) Nicoli, F.; Barth, A.; Bae, W.; Neukirchinger, F.; Crevenna, A. H.; Lamb, D. C.; Liedl, T. Directional Photonic Wire Mediated by Homo-Förster Resonance Energy Transfer on a DNA Origami Platform. *ACS Nano* **2017**.

(23) Klein, W. P.; Rolczynski, B. S.; Oliver, S. M.; Zadegan, R.; Buckhout-White, S.; Ancona, M. G.; Cunningham, P. D.; Melinger, J. S.; Vora, P. M.; Kuang, W.; et al. DNA Origami Chromophore Scaffold Exploiting HomoFRET Energy Transport to Create Molecular Photonic Wires. *ACS Appl. Nano Mater.* **2020**, *3* (4), 3323–3336.

(24) Bui, H.; Díaz, S. A.; Fontana, J.; Chiriboga, M.; Veneziano, R.; Medintz, I. L. Utilizing the Organizational Power of DNA Scaffolds for New Nanophotonic Applications. *Adv. Opt. Mater.* **2019**, *7* (18).

(25) Spillmann, C. M.; Stewart, M. H.; Susumu, K.; Medintz, I. L. Combining Semiconductor Quantum Dots and Bioscaffolds into Nanoscale Energy Transfer Devices. *Appl. Opt.* **2015**, *54* (31), F85.

(26) Hemmig, E. A.; Creatore, C.; Wünsch, B.; Hecker, L.; Mair, P.; Parker, M. A.; Emmott, S.; Tinnefeld, P.; Keyser, U. F.; Chin, A. W. Programming Light-Harvesting Efficiency Using DNA Origami. *Nano Lett.* **2016**, *16* (4), 2369–2374.

(27) Prinz, J.; Schreiber, B.; Olejko, L.; Oertel, J.; Rackwitz, J.; Keller, A.; Bald, I. DNA Origami Substrates for Highly Sensitive Surface-Enhanced Raman Scattering. *J. Phys. Chem. Lett.* **2013**, *4* (23), 4140–4145.

(28) Dutta, P. K.; Varghese, R.; Nangreave, J.; Lin, S.; Yan, H.; Liu, Y. DNA-Directed Artificial Light-Harvesting Antenna. *J. Am. Chem. Soc.* **2011**, *133* (31), 11985–11993.

(29) Albinsson, B.; Hannestad, J. K.; Börjesson, K. Functionalized DNA Nanostructures for Light Harvesting and Charge Separation. *Coord. Chem. Rev.* **2012**, *256* (21–22), 2399–2413.

(30) Olejko, L.; Bald, I. FRET Efficiency and Antenna Effect in Multi-Color DNA Origami-Based Light Harvesting Systems. *RSC Adv.* **2017**, *7* (39), 23924–23934.

(31) Kuzyk, A.; Jungmann, R.; Acuna, G. P.; Liu, N. DNA Origami Route for Nanophotonics. *ACS Photonics* **2018**, *5* (4), 1151–1163.

(32) Kaminska, I.; Bohlen, J.; Mackowski, S.; Tinnefeld, P.; Acuna, G. P. Strong Plasmonic Enhancement of a Single Peridinin–Chlorophyll a –Protein Complex on DNA Origami-Based Optical Antennas. *ACS Nano* **2018**, *12* (2), 1650–1655.

(33) Sasmal, D. K.; Pulido, L. E.; Kasal, S.; Huang, J. Single-Molecule Fluorescence Resonance Energy Transfer in Molecular Biology. *Nanoscale* **2016**, *8* (48), 19928–19944.

(34) Mathur, D.; Kim, Y. C.; Díaz, S. A.; Cunningham, P. D.; Rolczynski, B. S.; Ancona, M. G.; Medintz, I. L.; Melinger, J. S. Can a DNA Origami Structure Constrain the Position and Orientation of an Attached Dye Molecule? *J. Phys. Chem. C* **2021**, *125* (2), 1509–1522.

(35) Gopinath, A.; Thachuk, C.; Mitskovets, A.; Atwater, H. A.; Kirkpatrick, D.; Rothemund, P. W. K. Absolute and Arbitrary Orientation of Single-Molecule Shapes. *Science (80-. ).* **2021**, *371* (6531).

(36) Stennett, E. M. S.; Ma, N.; van der Vaart, A.; Levitus, M. Photophysical and Dynamical Properties of Doubly Linked Cy3-DNA Constructs. *J. Phys. Chem. B* **2014**, *118* (1), 152–163.

(37) Ranjit, S.; Gurunathan, K.; Levitus, M. Photophysics of Backbone Fluorescent DNA Modifications: Reducing Uncertainties in FRET. *J. Phys. Chem. B* **2009**.

(38) Monneret, S.; Bertaux, N.; Savatier, J.; Shaban, H. A.; Mavrakis, M.; Valades Cruz, C. A.; Kress, A.; Brasselet, S. Quantitative Nanoscale Imaging of Orientational Order in Biological Filaments by Polarized Superresolution Microscopy. *Proc. Natl. Acad. Sci. U. S. A.* **2016**.

(39) Rimoli, C. V.; Valades-Cruz, C. A.; Curcio, V.; Mavrakis, M.; Brasselet, S. 4polar-STORM Polarized Super-Resolution Imaging of Actin Filament Organization in Cells. *Nat. Commun.* **2022**, *13* (1), 301.

(40) Jungmann, R.; Steinhauer, C.; Scheible, M.; Kuzyk, A.; Tinnefeld, P.; Simmel, F. C. Single-Molecule Kinetics and Super-Resolution Microscopy by Fluorescence Imaging of Transient Binding on DNA Origami. *Nano Lett.* **2010**, *10* (11), 4756–4761.

(41) Hübner, K.; Joshi, H.; Aksimentiev, A.; Stefani, F. D.; Tinnefeld, P.; Acuna, G. P. Determining the In-Plane Orientation and Binding Mode of Single Fluorescent Dyes in DNA Origami Structures. *ACS Nano* **2021**, *15* (3), 5109–5117.

(42) Cervantes-Salguero, K.; Biaggne, A.; Youngsman, J. M.; Ward, B. M.; Kim, Y. C.; Li, L.; Hall, J. A.; Knowlton, W. B.; Graugnard, E.; Kuang, W. Strategies for Controlling the Spatial Orientation of Single Molecules Tethered on DNA Origami Templates Physisorbed on Glass Substrates: Intercalation and Stretching. *International Journal of Molecular Sciences*. 2022.

(43) Adamczyk, A. K.; Huijben, T. A. P. M.; Sison, M.; Di Luca, A.; Chiarelli, G.; Vanni, S.; Brasselet, S.; Mortensen, K. I.; Stefani, F. D.; Pilo-Pais, M.; et al. DNA Self-Assembly of Single Molecules with Deterministic Position and Orientation. *ACS Nano*

**2022**, *16* (10), 16924–16931.

(44) Enderlein, J. Modification of Förster Resonance Energy Transfer Efficiency at Interfaces. *Int. J. Mol. Sci.* **2012**, *13* (11), 15227–15240.

(45) Bohlen, J.; Cuartero-González, Á.; Pibiri, E.; Ruhlandt, D.; Fernández-Domínguez, A. I.; Tinnefeld, P.; Acuna, G. P. Plasmon-Assisted Förster Resonance Energy Transfer at the Single-Molecule Level in the Moderate Quenching Regime. *Nanoscale* **2019**, *11* (16), 7674–7681.

(46) Schuler, B. Single-Molecule FRET of Protein Structure and Dynamics - a Primer. *J. Nanobiotechnology* **2013**, *11* (1), S2.

(47) Zaza, C.; Chiarelli, G.; Zweifel, L. P.; Pilo-Pais, M.; Sisamakis, E.; Barachati, F.; Stefani, F. D.; Acuna, G. P. Super-Resolved FRET Imaging by Confocal Fluorescence-Lifetime Single-Molecule Localization Microscopy. *Small Methods* **2023**, *7* (7), 2201565.

(48) Agam, G.; Gebhardt, C.; Popara, M.; Mächtel, R.; Folz, J.; Ambrose, B.; Chamachi, N.; Chung, S. Y.; Craggs, T. D.; de Boer, M.; et al. Reliability and Accuracy of Single-Molecule FRET Studies for Characterization of Structural Dynamics and Distances in Proteins. *Nat. Methods* **2023**, *20* (4), 523–535.

(49) Introduction to Fluorescence BT - Principles of Fluorescence Spectroscopy; Lakowicz, J. R., Ed.; Springer US: Boston, MA, 2006; pp 1–26.